\documentclass[11pt]{article}
\usepackage[utf8]{inputenc}

\usepackage[T1]{fontenc}
\usepackage[english]{babel}
\usepackage[babel]{csquotes}
\usepackage{subfigure}
\usepackage{float}
\usepackage{graphicx}
\usepackage{amsmath}
\usepackage{amssymb}
\usepackage{amsfonts}
\usepackage{color}
\usepackage{braket}
\usepackage[bottom,hang,flushmargin]{footmisc} 
\usepackage{verbatim}
\usepackage{tensor}
\usepackage{bbold}
\usepackage{changepage}
\usepackage[
sorting=none
]{biblatex}
\usepackage[a4paper,left=2.5cm,right=2.5cm,top=2.5cm,bottom=2.5cm]{geometry}
\usepackage{scrextend}
\usepackage{pdfpages}
\usepackage{hhline}
\usepackage{pdflscape}
\usepackage{rotating}
\usepackage{tablefootnote}
\usepackage{tensor}
\usepackage{setspace}

\usepackage[hidelinks]{hyperref}

\newcommand{\bras}[1]{\tensor[]{\bra{#1}}{}}
\newcommand{\kets}[1]{\tensor[]{\ket{#1}}{}}

\makeatletter
\pretocmd{\section}{\addtocontents{toc}{\protect\addvspace{-35\p@}}}{}{}
\pretocmd{\subsection}{\addtocontents{toc}{\protect\addvspace{-11.3\p@}}}{}{}
\pretocmd{\subsubsection}{\addtocontents{toc}{\protect\addvspace{-12\p@}}}{}{}
\makeatother

 \def\d{\delta} 

\def\l{\lambda}

\def\cA{{\cal A}}  \def\cC{{\cal C}} 
   
 \def\cH{{\cal H}}  
  \def\cL{{\cal L}} 
\def\cM{{\cal M}}   
 \def\cQ{{\cal Q}}

\def\R{{\mathbb R}}  

 \def\one{\mbox{1 \kern-.59em {\rm l}}}

\newcommand{\Tr}{\mathrm{Tr}}

\newcommand{\End}{\mathrm{End}}

\title{Master Thesis - Paper}
\author{Laura Felder}
\date{March 2023}

\begin{document}

\begin{flushright}
 UWThPh-2023-21 \\
\end{flushright}
\hfill
\vskip 0.01\textheight

\begin{center}
{ \Large\textbf{Oxidation, Reduction and Semi-Classical Limit for  \\ Quantum Matrix Geometries}}

{ \Large Laura O. Felder\footnote{\href{mailto:a11705898@unet.univie.ac.at}{a11705898@unet.univie.ac.at}}, Harold C. Steinacker\footnote{\href{mailto:harold.steinacker@univie.ac.at}{harold.steinacker@univie.ac.at}}}

\textit{Faculty of Physics, University of Vienna\\
Boltzmanngasse 5, A-1090 Vienna, Austria}
\end{center}

\begin{abstract}
    Matrix configurations define noncommutative spaces endowed with extra structure including a generalized Laplace operator, and hence  a metric structure. Made dynamical via matrix models, they describe rich physical systems including noncommutative gauge theory and emergent gravity.
    Refining the construction in \cite{Steinacker_2021},     
    we construct a semi-classical limit through  an immersed submanifold of complex projective space based on quasi-coherent states.
    We observe the phenomenon of oxidation, where 
   the resulting semi-classical space acquires spurious extra dimensions. We propose to remove this artifact by
    passing to a leaf of a carefully chosen foliation,
   which allows to extract the geometrical content of the noncommutative spaces. This is demonstrated numerically via multiple examples. 
\end{abstract}

\section{Introduction}

The idea of noncommutative geometry is to extend geometric notions beyond the realm of classical manifolds, working with noncommutative algebras of functions in the spirit of quantum mechanics. A specific realization of this idea is provided by
matrix models in the context of string theory \cite{Ishibashi_1997,Banks:1996vh}, which describe 
noncommutative branes. 
More specifically, a number of standard solutions can be interpreted as quantized symplectic spaces, which provides a large  class of backgrounds, which is expected to be closed under small perturbations.

This leads to the problem of extracting such an underlying geometry from a given matrix configuration. This is an interesting and meaningful problem at least for \textit{almost-commutative} matrix configurations, and a first systematic framework to address this problem was developed in \cite{Steinacker_2016}, 
based on previous ideas involving (quasi-)coherent states \cite{Ishiki:2015saa,Perelomov_1986}.
In particular, a concept of a quantum space or manifold was introduced, which can be associated to any given matrix configuration.

However, this approach is not yet completely satisfactory, as this quantum manifold often turns out to 
be \textit{oxidized} with spurious extra dimensions. 
Then the proper or minimal underlying
geometry is not easily recognized.
This leads to the problem of explicitly finding 
such a minimal, semi-classical description for generic, almost-commutative matrix configurations.
This is the motivation for the present work. We investigate several possible strategies to achieve this goal, which are tested numerically in 
non-trivial examples.
In particular, we can offer a method to determine such an appropriate semi-classical description, by considering  certain leaves defined by the quantum metric or would-be symplectic form, which is tested numerically.
We also provide some new and refined results on the underlying geometrical structures, including a refined concept of a quantum manifold and a proof of its manifold structure following \cite{Felder_2023}.

To put the present work into context, we  note that 
a somewhat simpler approach to extract the underlying geometry using quasi-coherent states was given in \cite{Schneiderbauer_2015,Schneiderbauer_2016}, based on previous ideas \cite{Berenstein:2012ts}  in the context of string theory. Some 
 of the theoretical concepts were  also considered previously in \cite{Ishiki:2015saa} in a somewhat different way.
The numerical investigations of the present work are based on the Mathematica package \textit{QMGsurvey} \cite{Felder_2023_Software,Felder_2023}
developed by one of the authors.

The content of this paper is organized as follows.
Section \ref{QMG} reviews the definition of the quasi-coherent states and the quantum manifold, mostly based on \cite{Steinacker_2021} and \cite{Felder_2023}. In particular, the concept of quantization maps for Poisson manifolds is reviewed in section \ref{QS}. Section \ref{MCQCS} introduces matrix configurations and quasi-coherent states associated to them. The hermitian form and the null spaces are discussed in section \ref{herm} and \ref{NullSpace} respectively. Then, in section \ref{mfp} the crucial quantum manifold and its partner, the embedded quantum space, are defined. In order to make the interpretation as a semi-classical limit manifest, a preliminary quantization map is introduced in section \ref{limPrelim}.\\
In section \ref{LEAVES}, we discuss foliations of this quantum manifold. Section \ref{leavesGeneral} formulates some requirements for the resulting leaves,  and in particular  the \textit{hybrid leaf} is introduced
in section \ref{hybridLeaf}, which appears to be most promising \cite{Felder_2023}. Finally, in section \ref{lim} a refined quantization map is introduced.\\
We then study a number of 
examples in section \ref{EXAMPLES}, with particular focus on the squashed fuzzy sphere (section \ref{sfs}). The more generic perturbed fuzzy spheres (section \ref{pfs}) can be interpreted in terms of gauge fields. The fuzzy torus (section \ref{ft}) is an example which is not obtained from a quantized coadjoint orbit, while the squashed fuzzy $\mathbb{C}P^2$ (section \ref{sfcp2}) provides a higher-dimensional example.\\
Finally in appendix \ref{appendix:analyticity}, the analyticity and especially the smoothness of the quasi-coherent states are discussed, while appendix \ref{appendix:quantummanifold} features a  proof that the quantum manifold is a smooth manifold based on \cite{Felder_2023}.

\section{Quantum Matrix Geometries and Quasi-Coherent States}
\label{QMG}

In this section, we provide a concise discussion of  matrix configurations and their associated structures, with the aim to extract the underlying geometry explicitly. 
  This leads to a refinement of the basic definitions and results in \cite{Steinacker_2021}, and in particular to a more general and refined proof of the manifold structure of the quantum manifold and the embedded quantum space in section \ref{mfp}.
  We discuss the concepts of oxidation and reduction, which arises in the practical approaches to extract the underlying classical geometry from a given matrix configuration. For a more thorough discussion and more examples see \cite{Felder_2023}.

\subsection{Quantum Spaces and Quantization Maps}
\label{QS}

Given a Poisson manifold (a manifold $\mathcal{M}$ endowed with a Poisson bracket  $\{,\}:\mathcal{C}^\infty(\mathcal{M})\times\mathcal{C}^\infty(\mathcal{M})\to\mathcal{C}^\infty(\mathcal{M})$ which satisfies the Leibniz rule and the Jacobi identity), we may consider its Poisson algebra of smooth complex valued functions $\mathcal{C}(\mathcal{M}):=\mathcal{C}^\infty(\mathcal{M},\mathbb{C})$.
This is of course a commutative algebra, describing the underlying 
\textit{commutative} or \textit{classical space} $\cM$.\\
In noncommutative geometry\footnote{There are various concepts of noncommutative geometry in the literature. Here we  follow a pragmatic, physics-oriented approach based on matrices and matrix models, rather than  an axiomatic approach.}, the commutative algebra $\mathcal{C}(\mathcal{M})$
is replaced by the noncommutative algebra of endomorphisms $\operatorname{End}(\mathcal{H})$ 
in some (separable) Hilbert space $\mathcal{H}$. Amended with extra structure, this is used to describe
a \textit{noncommutative space} respectively a \textit{quantum space}. Then the commutator $[,]$ 
 naturally replaces the Poisson bracket, as it also fulfills the Jacobi identity and a generalized Leibniz rule. The structural correspondence between these classical and quantum concepts 
is compared in table \ref{compareCQ}.

\begin{table}[H]
\begin{tabular}{c|c|c}
\textbf{structure} & \textbf{classical space} & \textbf{quantum space} \\ \hline
algebra & $\mathcal{C}(\mathcal{M})$ & $\operatorname{End}(\mathcal{H})$ \\
addition \& multiplication & pointwise operations & matrix operations \\
(Lie) bracket & $(f,g)\mapsto i\{f,g\}$ & $(F,G)\mapsto[F,G]$ \\
conjugation & $f\mapsto f^*$ & $F\mapsto F^\dagger$ \\
inner product\tablefootnote{$\operatorname{dim}(\mathcal{M})=:2n$ and $\Omega:=\frac{1}{n!}\omega^{\wedge n}=\sqrt{\operatorname{det}(\omega_{ab})}dx^1\wedge\dots\wedge dx^{2n}\in\Omega^{2n}(\mathcal{M})$ is the volume form arising from the symplectic form $\omega$ defined by $\{,\}$ in the non-degenerate case.} (if nondeg.) & $(f,g)\mapsto\braket{f\vert g}_2:=\frac{1}{(2\pi)^n}\int_\mathcal{M}\Omega\, f^*g$ & $(F,G)\mapsto\braket{F\vert G}_{HS}:=\operatorname{tr}(F^\dagger G)$ \\
observable & $f^*=f$ & $F^\dagger=F$ \\
mixed state & $f\geq0$ \& $\Vert f\Vert_2=1$ & $F\geq0$ \& $\Vert F\Vert_{HS}=1$\\
\end{tabular}
\centering
\caption{Comparison of related structures on Poisson manifolds and endomorphism algebras of Hilbert spaces, for $f,g\in\mathcal{C}(\mathcal{M})$ and $F,G\in \operatorname{End}(\mathcal{H})$.}
\label{compareCQ}
\end{table}

Accordingly, to \textit{quantize} a classical space $\cM$ we should replace any element of $\mathcal{C}(\mathcal{M})$ by an element of $\operatorname{End}(\mathcal{H})$. This is formalized in terms of a linear map called \textit{quantization map}
\begin{align}
    \cQ:\quad \mathcal{C}(\mathcal{M})\to\operatorname{End}(\mathcal{H})\ ,
\end{align}
 depending on a \textit{quantization parameter}\footnote{One should think of it as a formalization of $\hbar$.} $\theta$, which
satisfies the following axioms:
\vspace{-0.5cm}
\begin{enumerate}
    \item $\cQ(1_\mathcal{M})=\mathbb{1}_\mathcal{H}$ (completeness relation)
    \item $\cQ(f^*)=\cQ(f)^\dagger$ (compatibility of con- and adjugation)
    \item $\lim_{\theta\to0} (\cQ(f\cdot g)-\cQ(f)\cdot \cQ(g))=0\quad\text{and}\quad\lim_{\theta\to0}\frac{1}{\theta} (\cQ(\{f, g\}))-\frac{1}{i} [\cQ(f),\cQ(g)])=0$ \\(asymptotic compatibility of algebra structure)
    \item $[\cQ(f),F]=0\quad\forall f\in \mathcal{C}(\mathcal{M})\implies F\propto\mathbb{1}_\mathcal{H}$ (irreducibility) \ .
\end{enumerate}

There are two optional additional axioms which in many cases make the choice of a quantization map unique:
\vspace{-0.5cm}
\begin{enumerate}
\setcounter{enumi}{4}
    \item $\braket{\cQ(f)\vert \cQ(g)}_{HS}=\braket{f\vert g}_2$ (isometry)
    \item $\cQ(g\triangleright f)=g\triangleright \cQ(f)\quad \forall g\in G$ (intertwiner of group action) \ .
\end{enumerate}
\vspace{-0.5cm}
Here we assume that both $\mathcal{C}(\mathcal{M})$ and $\operatorname{End}(\mathcal{H})$ are endowed with a group action denoted by $\triangleright$ of some given Lie group $G$ \cite{Steinacker_2016II,Schneiderbauer_2016,Waldmann_2007}.

\subsection{Matrix Configurations and Quantum Matrix Geometries}

We briefly recall some basic concepts following \cite{Steinacker_2021}.
A {\bf \em matrix configuration} 
is a collection of $D$ 
hermitian matrices $X^a \in \End(\cH)$ for $a=1,...,D$, where
$\cH$ is a finite-dimensional Hilbert space. 
We will focus on irreducible matrix configurations, which means that
the only matrix which commutes with all $X^a$ is the identity matrix.
For any such matrix configuration, we consider an action defined by a matrix model with the structure
\begin{align}
 S[X] = \Tr([X^a,X^b][X_a,X_b] + m^2 X^a X_a)
 \label{MM-Euclid-mass}
\end{align}
or similar. Here 
indices are contracted with $\d^{ab}$, which is interpreted as Euclidean metric on target space $(\R^D,\d)$.

For a random matrix configuration $X^a$, one should not expect to find a reasonable geometric interpretation.
However, the above action selects matrix configurations where all commutators $[X^a,X^b]$ are small; 
such matrix configurations are denoted as \textit{almost commutative}. It then make sense to interpret the commutators as quantized Poisson brackets, in the spirit of section \ref{QS}. 
More specifically,  we should expect that it can be described  by 
a quantized Poisson manifold $(\mathcal{M},\{,\})$, in the sense that 
\begin{align}
    X^a = \cQ(x^a)
    \label{quantiz-map-X}
\end{align}
where 
\begin{align}
    x^a:\quad \cM \hookrightarrow \R^D
\end{align}
is a (smooth) embedding of a Poisson manifold in target space $\R^D$. 
Our aim is to extract the underlying Poisson or symplectic manifold $\cM$ and its embedding in target space via $x^a$, such that \eqref{quantiz-map-X} holds at least in a suitable {\em semi-classical regime} of  IR modes with sufficiently large wavelength.
We shall denote these geometrical data as \textit{semi-classical limit} of the matrix configuration.
Conversely, a matrix configuration arising as quantization of a symplectic space embedded in $\R^D$ will be denoted as quantized brane.

We therefore want to address the following general problem: given some matrix configuration 
consisting of $D$ hermitian matrices $X^a,\ a=1,...,D$, is there a
symplectic manifold $\cM \subset \R^D$ embedded\footnote{Here ``embedding'' is understood in 
a loose sense; the embedding map may be degenerate.} in $\R^D$ via some  
map $x^a: \cM \hookrightarrow \R^D$  
such that the $X^a$ can be viewed in a meaningful way as quantization of classical embedding functions $x^a$, i.e. as a quantized brane? And if yes, how can we determine this manifold $\cM$ explicitly?

The above idea needs some sharpening to become meaningful. As explained in \cite{Steinacker_2021}, the relation 
between classical functions $\cC(\cM)$ and quantum ``functions'' $\End(\cH)$ given by $\cQ$ should be restricted to a small regime of IR modes, such that the restricted quantization map
\begin{align}
   L^2(\cM) \supset \cC_{IR}(\cM) \quad\stackrel{\cQ}{\to} \quad Loc(\cH) \subset \End(\cH)
\end{align}
is an (approximate) isometry,
 where 
$\End(\cH)$ is equipped with the Hilbert-Schmidt Norm. Therefore our task is to extract a semi-classical geometry which admits a sufficiently large regime of IR modes such that $\cQ$ is an approximate isometry.

\subsection{Oxidation and Reduction of Quantum Matrix Geometries}
\label{sec:reduction}

Assume we are given some almost-commutative matrix configuration $\bar X^a \in \End(\cH)$, which can be interpreted as quantized embedded symplectic space $\bar\cM\subset\R^D$ in the above sense. 
Then consider some deformation of it given by 
\begin{align}
 X^a = \bar X^a + \cA^a(\bar X) \
 \label{oxidation}
\end{align}
where $\cA^a(\bar X) \ \in \End(\cH)$ is some function of the $\bar X^a$. As
long as the $\cA^a$ are sufficiently mild, this deformed
matrix configuration should clearly be interpreted as deformed quantized embedding of the same underlying $\bar\cM$ in target space $\R^D$:
\begin{align}
 x^a = \bar x^a + \cA^a(\bar x): \quad \bar\cM \hookrightarrow \R^D \ .
\end{align}
However, the \textit{abstract quantum space} $\cM$ (as defined below)   defined by
$X^a$ may  have larger dimension than $\bar\cM$, as
$\bar\cM$ grows some ``thickness'' in transverse directions. This is a spurious and undesirable effect denoted as \textit{oxidation}.
Such a situation is not easily recognized in terms of the procedure decribed below, and it would be desirable
to extract the underlying ``reduced''  structure of $\bar\cM$. This problem is one of the motivations of the present paper.

Various methods to remove the oxidation are conceivable, such as looking for
 minima of the lowest eigenvalue function $\l(x)$, or identifying a hierarchy
of the eigenvalues of $\omega$ and $g$.
Another strategy may be to impose the quantum K\"ahler condition,
which may hold only on some sub-variety of the abstract quantum space $\cM$.
In any case, it would be very desirable
to have efficient tools and algorithms to numerically ``measure'' and 
determine the underlying quantum space corresponding to some generic
matrix configuration; for first steps towards such a goal see e.g.
\cite{Schneiderbauer_2016}.

\paragraph{Effective metric and physical significance}

 Assuming that we have identified a given matrix configuration with a quantized embedded symplectic space embedded in target space $\cM\hookrightarrow \R^D$, it is automatically equipped with rich geometrical structure including an induced metric. However, it turns out that the propagation of fluctuations on such a matrix background in Yang-Mills matrix models is governed by a different {\em effective} metric. Although this is not the main subject of this paper, this issue deserves some discussion:

 The effective metric can be understood as follows:
 Every element $D\in\operatorname{End}(\mathcal{H})$ defines a derivation\footnote{That is, a linear map $\widehat{D}:\operatorname{End}(\mathcal{H})\to\operatorname{End}(\mathcal{H})$ that satisfies $\widehat{D}(F\cdot G)=F\cdot \widehat{D}(G)+\widehat{D}(F)\cdot G\;\forall F,G\in\operatorname{End}(\mathcal{H})$.} on $\operatorname{End}(\mathcal{H})$ via $\widehat{D}:=[D,\cdot]$, considered as a first order differential operator. Then the map
\begin{align}
    \square: \quad &\operatorname{End}(\mathcal{H})\to \operatorname{End}(\mathcal{H})\nonumber\\
    & F\mapsto \square F:= [X^a,[X_a,F]]
\end{align}
defines the  \textit{matrix Laplacian}, which should be interpreted as a second order differential operator on $\operatorname{End}(\mathcal{H})$. This operator encodes the information of a metric \cite{Steinacker_2021,Steinacker_2016II,Schneiderbauer_2016}. Here indices are raised and lowered with the ``target space metric'' $\d_{ab}$ on $\R^D$.

\subsection{Matrix Configurations and Quasi-Coherent States}
\label{MCQCS}

Our aim is to construct a Poisson manifold $(\mathcal{M},\{,\})$ -- or preferrably a symplectic manifold\footnote{Note that every symplectic manifold carries a non-degenerate Poisson bracket, whereas every Poisson manifold decays into a foliation of symplectic leaves \cite{Michor_2008}.} $(\mathcal{M},\omega)$ -- embedded into Euclidean space via $x^a:\ \mathcal{M}\hookrightarrow \mathbb{R^D}, a=1,...,D$ together with a quantization map\footnote{The role of the quantization parameter $\theta$ is not evident here. In some examples (such as fuzzy spheres), one may consider families of matrix configurations  parameterized by $N$ and consider $\theta:=\frac{1}{N}$. However in general, this notion is replaced by the choice of an appropriate semi-classical regime.} $\cQ:\mathcal{C}(\mathcal{M})\to\operatorname{End}(\mathcal{H})$,
such that
\begin{align}
    \cQ(x^a)= X^a \ .
\end{align}

We follow the approach proposed in \cite{Steinacker_2021}  based on so called \textit{quasi-coherent states}, generalizing the coherent states on the Moyal-Weyl quantum plane.
Therefore, we introduce the \textit{Hamiltonian}
\begin{align}
    \label{Hamiltonian}
     H:\quad &\mathbb{R}^D\to\operatorname{End}(\mathcal{H})\nonumber\\
    &\left(x^a\right)\mapsto H_x:=\frac{1}{2}\sum_a\left(X^a-x^a\mathbb{1}\right)^2\ ,
\end{align}
defining a positive definite\footnote{Definiteness follows readily from irreducibility of the matrix configuration \cite{Steinacker_2021}.} hermitian operator $H_x$ at every point $x=(x^a)\in\mathbb{R}^D$. In analogy to string theory, we call $\mathbb{R}^D$ \textit{target space}.

Thus, for every point $x$ in target space, there is a lowest eigenvalue $\lambda(x)$ of $H_x$ with corresponding eigenspace $E_x$. Based on that, we define the set
\begin{align}
    \Tilde{\mathbb{R}}^D:=\{x\in \mathbb{R}^D\vert \operatorname{dim}(E_x)=1\} \ .
\end{align}
This restriction may look artificial at first, but it will be essential to obtain the appropriate topology.
In particular, we can then choose 
for every point $x\in\Tilde{\mathbb{R}}^D$ a normalized vector $\ket{x}\in E_x$ which is unique up to a $U(1)$ phase.
Such a vector is called \textit{quasi-coherent state} (at $x$) \cite{Steinacker_2021}.
\\
Sometimes we will need the whole eigensystem of $H_x$, denoted as\footnote{Possible ambiguity will not be important.}
$H_x\ket{k,x}=\lambda^k(x)\ket{k,x}$ and ordered such that  $\lambda^k(x)\leq\lambda^l(x)$ for $k<l$.

\subsection{The Hermitian Form}
\label{herm}

As explained in appendix\footnote{In particular, $\Tilde{\mathbb{R}}^D\subset\mathbb{R}^D$ is always open.} \ref{appendix:analyticity},
we can locally choose the quasi-coherent states depending on $x \in \Tilde{\mathbb{R}}^D$ in a smooth way;
this will be assumed in the following. 
We can then define
\begin{align}
    iA_a(x):=\bras{x}\partial_a\kets{x}
\end{align}
and 
\begin{align}
    D_a\kets{x}:=(\partial_a-iA_a)\kets{x}=(\operatorname{1}-\kets{x}\bras{x})\partial_a\kets{x}\ ,
\end{align}
where $A_a$ has the interpretation of a connection 1-form and $D_a$ of a covariant derivative, observing that $D_a\kets{x}\mapsto e^{i\phi(x)}D_a\kets{x}$ 
under a local $U(1)$ transformation $\kets{x}\mapsto e^{i\phi(x)}\kets{x}$ for any $\phi\in\mathcal{C}^\infty(\mathcal{M})$.
Consequently, the \textit{hermitian form}
\begin{align}
    h_{ab}(x):=(D_a\kets{x})^\dagger D_a\kets{x}=:\frac{1}{2}(g_{ab}(x)-i\omega_{ab}(x))
\end{align}
is invariant under local $U(1)$ transformations. The real symmetric object $g_{ab}$ is called \textit{quantum metric} (for its interpretation see \cite{Steinacker_2021}; it should not be confused with the effective metric, which is encoded e.g. in $\Box$), and the real antisymmetric form $\omega_{ab}$ is called \textit{(would-be) symplectic form}, as
\begin{align}
    \omega_{ab}=\partial_aA_b-\partial_bA_a=(dA)_{ab} \ 
\end{align}
is closed and hence  defines a symplectic form if  nondegenerate.
These objects will play an important role, and we will meet them again in section \ref{qmfld}.
\\
For later use, we also define two further concepts: the \textit{embedded point} (at $x$)
\begin{align}
    \mathbf{x}^a(x):=\bra{x}X^a\ket{x}, \quad \mathbf{x}(x)=(\mathbf{x}^a(x))
\end{align}
and the \textit{(would-be) induced Poisson tensor} (at $x$)
\begin{align}
    \theta^{ab}(x):=\frac{1}{i}\bra{x}[X^a,X^b]\ket{x} \ .
\end{align}
Exploiting the simple structure of $H_x$, one shows that
\begin{align}
    (\partial_a-iA_a)\kets{x}=\mathfrak{X}_x^a\ket{x}
\end{align}
where
\begin{align}
\mathfrak{X}^a_x:=\sum_{k=2}^N  \frac{\ket{k,x}\bra{k,x}}{\lambda^k(x)-\lambda(x)}X^a
 = (H_x-\lambda(x))^{-1'} X^a
\end{align}
Here 
\begin{align}
    \sum_{k=2}^N  \frac{\ket{k,x}\bra{k,x}}{\lambda^k(x)-\lambda(x)} = (H_x-\lambda(x))^{-1'}
\end{align}
is the \textit{pseudo inverse} of $H_x-\lambda(x) \mathbb{1}$.
We obtain the remarkable result
\begin{align}
    h_{ab}(x)=\bra{x}(\mathfrak{X}^a_x)^\dagger \mathfrak{X}^b_x\ket{x} \ .
\end{align}
which  allows us to calculate $h_{ab}$ without having to perform any derivatives; this  is clearly desirable in numerical calculations.
Similarly, we also observe that
\begin{align}
    (\partial_a\mathbf{x}^a)(x)= \bra{x}X^a\mathfrak{X}^b\ket{x}+\bra{x}X^b\mathfrak{X}^a\ket{x}
\end{align}
which is also helpful in some numerical calculations \cite{Steinacker_2021, Felder_2023}.

\subsection{The Null Spaces}
\label{NullSpace}

The explicit form of $H_x$ can be used to establish some global properties of the quasi-coherent states. One easily verifies
\begin{align}
    H_x=H_y+\frac{1}{2}\left(\vert x\vert^2-\vert y\vert^2\right)\mathbb{1}-\sum_a(x^a-y^a)X^a \label{HPerturbation}
\end{align}
and
\begin{align}
    \label{HStraightLines}
    H_{(1-\alpha)x+\alpha y}=(1-\alpha)H_x+\alpha H_y+\frac{\alpha^2-\alpha}{2}\vert x-y\vert^2\mathbb{1}\quad \forall\alpha\in\mathbb{R} \ .
\end{align}
It may occur that some of the lowest eigenspaces coincide for different points in $\Tilde{\mathbb{R}}^D$, defining an equivalence relation in $\Tilde{\mathbb{R}}^D$ by 
\begin{align}
   x\sim y \ \iff E_x=E_y \ .
\end{align}
We denote the resulting equivalence classes 
$\mathcal{N}_x:=[x]_\sim=\{y\in\Tilde{\mathbb{R}}^D\vert E_x=E_y\}$ \textit{null space} (of $x$). These spaces turn out to be very important to the following construction.

Assume that $x\sim y$. Then from equation (\ref{HStraightLines}) it follows directly that $\ket{x}$ is an eigenstate of $H_{(1-\alpha)x+\alpha y}$ of eigenvalue $(1-\alpha)\lambda(x)+\alpha \lambda(y)+\frac{\alpha^2-\alpha}{2}\vert x-y\vert^2$ and it is obvious that for $\alpha\in[0,1]$ this is the lowest eigenspace. But then it is immediate that for $\alpha\in[0,1]$ we have $(1-\alpha)x+\alpha y\in\mathcal{N}_x$. Thus
\begin{align}
    \mathcal{N}_x \text{ is convex}\ .
\end{align}
Further, the line $(1-\alpha)x+\alpha y$ only ceases to be in $\mathcal{N}_x$ if $E_x$ stops to be the unique lowest eigenspace of $H_x$, thus at a point in the complement of $\Tilde{\mathbb{R}}^D$.
This then implies that
\begin{align}
    \mathcal{N}_x\subset\Tilde{\mathbb{R}}^D \text{ is closed} \ .
\end{align}
Assume again that $x\sim y$. Using equation (\ref{HPerturbation}), it follows that $\sum_a (x-y)^a X^a\ket{x}=(\lambda(y)-\lambda(x)+\frac{1}{2}(\vert x\vert^2-\vert y\vert^2)) \ket{x}$ and acting with $(H_x-\lambda(x))^{-1'}$  from the left, we find $\sum_a(x-y)^a\mathfrak{X}^a\ket{x}=0$ \cite{Steinacker_2021}.
\\
The previous results imply that $\mathcal{N}_x$ is a submanifold of $\Tilde{\mathbb{R}}^D$, and we can consider $v\in T_x\mathcal{N}_x\subset T_x\mathbb{R}$. Of course $v\propto(x-y)$ for some $y\in \mathcal{N}_x$ and consequently
\begin{align}
    T_x\mathcal{N}_x\subset\operatorname{ker}(h_{ab}) \ .
\end{align}

\subsection{The Quantum Manifold}
\label{qmfld}
\label{mfp}

The main step in the construction of a semi-classical limit is the definition of the \textit{quantum manifold}.
Following \cite{Steinacker_2021}, we define the collection of all quasi-coherent states modulo $U(1)$ as
\begin{align}
    \mathcal{M}':=\cup_{x\in\Tilde{\mathbb{R}}^D}U(1)\ket{x}/U(1) 
   \cong\left\{E_x\vert x\in\Tilde{\mathbb{R}}^D\right\}\subset\mathbb{C}P^{N-1} \ ,
   \label{qmPrime}
\end{align}
which is a subspace of complex projective space (by identifying $\mathcal{H}\cong\mathbb{C}^N$).

Locally, every $\kets{\cdot}$ defines a smooth map $q_s:=U(1)\kets{\cdot}:U\subset\Tilde{\mathbb{R}}^D\to\mathcal{M}'$.
If we want to be precise, we can use the canonical smooth projection $p:\mathbb{C}^N\to\mathbb{C}P^{N-1}$ and identify $q_s\cong p\circ \kets{\cdot}$.
Since all maps $\kets{\cdot}$ only deviate in a $U(1)$ phase, all $q_s$ assemble to a unique smooth map $q:\Tilde{\mathbb{R}}^D\to\mathcal{M}'$.

Yet, in general, the tangent map $T_xq$ may not have constant rank, preventing $\mathcal{M}'$ from being a manifold. To address this issue
we define the \textit{maximal rank} $k:=\max_{x\in\Tilde{\mathbb{R}}^D}\operatorname{rank}(T_xq)$
and the sets
\begin{align}
    \widehat{\mathbb{R}}^D &:=\{x\in\Tilde{\mathbb{R}}^D\vert \operatorname{rank}(T_xq)=k\},\nonumber \\
\mathcal{M} &:=q(\widehat{\mathbb{R}}^D)\subset\mathbb{C}P^{N-1} \quad\text{and}\nonumber \\
\Tilde{\mathcal{M}} &:=\mathbf{x}(\widehat{\mathbb{R}}^D)\subset\mathbb{R}^D
\end{align}
where $\mathcal{M}$ is called \textit{(abstract) quantum manifold} and $\Tilde{\mathcal{M}}$ is the \textit{embedded quantum space} or \textit{brane}.
Note that $\widehat{\mathbb{R}}^D\subset\mathbb{R}^D$ is open\footnote{See for example the discussion of definition 2.1 in \cite{Michor_2008}.}.

Based on the results of section \ref{NullSpace} and the constant rank theorem, it is proven
in appendix \ref{appendix:quantummanifold}  that $\mathcal{M}$ is a smooth manifold of dimension $k$ immersed into $\mathbb{C}P^{N-1}$.

Thus, we can pull back the Fubini–Study metric and the Kirillov-Kostant-Souriau symplectic form along the immersion $\mathcal{M}\hookrightarrow\mathbb{C}P^{N-1}$ denoted as $g_\mathcal{M}$ (a Riemannian metric) and $\omega_\mathcal{M}$ (a closed 2-form).
It is shown
in appendix \ref{appendix:quantummanifold} that these exactly reproduce $g_{ab}$ and $\omega_{ab}$ if further pulled back along $q$ to $\widehat{\mathbb{R}}^D$.\\
We can conclude that the kernel of $g_{ab}(x)$ coincides with the kernel of $T_xq$ (since $g_\mathcal{M}$ is nondegenerate), but we only know that the kernel of $T_xq$ lies within the kernel of $\omega_{ab}(x)$ which may have a further degeneracy.

While we have nice results for the quantum manifold, we do not know much about the regularity of the embedded quantum space. Yet, we can certainly interpret the surjection 
\begin{align}
    \label{EmbeddingFunction}
    \mathbf{x}:\; & \mathcal{M}\to\Tilde{\mathcal{M}}\subset\mathbb{R}^D\nonumber\\
    &U(1)\ket{\psi}\mapsto (\bra{\psi}X^a\ket{\psi})
\end{align}
as \textit{(would-be) Cartesian embedding functions}, giving  geometrical meaning to the latter \cite{Steinacker_2021, Felder_2023}.

\subsection{The Preliminary Quantization Map}
\label{limPrelim}

Assuming that $\omega_\mathcal{M}$ is nondegenerate and hence a symplectic form, we can define the symplectic volume form $\Omega_\mathcal{M}:=\frac{1}{(k/2)!}\omega_\mathcal{M}^{\wedge k/2}$, allowing us to integrate over $\mathcal{M}$.\\
As the points of $\mathcal{M}$ are given by cossets $U(1)\ket{\psi}$ of quasi-coherent states, each point provides us with a unique projector $\ket{\psi}\bra{\psi}$. Slightly abusing the notation we write $\ket{\cdot}\bra{\cdot}$ for the map $U(1)\ket{\psi}\mapsto\ket{\psi}\bra{\psi}$.
We can then define the preliminary quantization map
\begin{align}
    \label{quantmapCS}
    \cQ:\quad &\mathcal{C}(\mathcal{M})\to\operatorname{End}(\mathcal{H})\nonumber\\
    &\phi\mapsto \frac{\alpha}{(2\pi)^{k/2}}\int_\mathcal{M}\Omega_\mathcal{M} \,\phi\ket{\cdot}\bra{\cdot} \ ,
\end{align}
where we choose $\alpha$ such that the trace of the completeness relation is satisfied, i.e.
\begin{align}
    \frac{\alpha}{(2\pi)^{k/2}}V_\omega=N
\end{align}
with the \textit{symplectic volume} $V_\omega :=\int_\mathcal{M}\Omega_\mathcal{M}$.
Similarly, we can define a
\textit{de-quantization map}
\begin{align}
    \operatorname{Symb}:\; &\operatorname{End}(\mathcal{H})\to\mathcal{C}(\mathcal{M})\nonumber\\
    &\Phi\mapsto \bra{\cdot}\Phi\ket{\cdot} \ ,
\end{align}
called \textit{symbol map}. Then, $\mathbf{x}^a$ is the \textit{symbol} of $X^a$.

Looking at specific examples, it turns out that the above construction of $(\mathcal{M},\omega_\mathcal{M})$ together with $\cQ:\mathcal{C}(\mathcal{M})\to\operatorname{End}(\mathcal{H})$ and $\mathbf{x}:\mathcal{M}\to\mathbb{R}^D$ does not always provide an adequate semi-classical description of  matrix configurations $(X^a)$.
This is essentially the issue of oxidation mentioned above. More specifically, the main problem is that $\omega_\mathcal{M}$ may be  degenerate, so that $\cQ$ is not well defined.
This and other problems are tackled in the next section, leading to a  refinement of the above construction \cite{Steinacker_2021, Felder_2023}.

\section{Foliations and Leaves of the Quantum Manifold}
\label{LEAVES}

As already stated, the most problematic issue of the preliminary quantization map (\ref{quantmapCS}) is that $\cQ$ is  the zero map if $\omega_\mathcal{M}$ is degenerate. 
This already happens in 
simple examples such as the squashed fuzzy sphere. Moreover,
the same example shows that the quantum manifold $\mathcal{M}$ is not stable under perturbations \eqref{oxidation}: while the maximal rank $k$ equals two for the round fuzzy sphere, it jumps to three for any infinitesimal squashing \cite{Felder_2023}. This phenomenon is denoted as \textit{oxidation}.
Similarly, the \textit{effective dimension} $l$ of $\mathcal{M}$ 
 may be significantly smaller than $k=\operatorname{dim}(\mathcal{M})$.
Here \textit{effective dimension} is a heuristic concept which may be defined e.g. in terms of the eigenvalues of the quantum metric $(g_{ab}(x))$:  Introducing a suitable cutoff $C$ in the presence of a natural gap in the spectrum of $g_{ab}(x)$, we can define the effective dimension $l$ as the number of eigenvalues larger than $C$.
It may then be appropriate  to replace the abstract quantum manifold with a simpler one with dimension $l$; this is denoted  by \textit{reduction}.
In the example of a moderately squashed fuzzy sphere, the appropriate effective dimension is clearly two, which is smaller than the maximal rank $k=3$.

In the following sections, we discuss some natural strategies  to achieve this reduction using the idea of leaves, and some variations thereof.
The resulting modified quantization map and semi-classical limit are described.

\subsection{Leaves of the Quantum Manifold}
\label{leavesGeneral}

The above problems may be addressed by passing from $\mathcal{M}$ to an appropriately chosen leaf $\mathcal{L}\subset \mathcal{M}$ of a foliation of $\mathcal{M}$.

Such a leaf should then satisfy the following conditions:
\vspace{-0.5cm}
\begin{itemize}
    \item $\omega_\mathcal{L}$ (the pullback of $\omega_\mathcal{M}$ to $\mathcal{L}$) should be nondegenerate, making it into a symplectic form. \\This implies that $\operatorname{dim}(\mathcal{L})$ has to be even.
    \item The dimension of $\mathcal{L}$ should agree with the effective dimension: $l=\operatorname{dim}(\mathcal{L})$.
    \item The leaf should contain the directions that are not suppressed by the quantum metric.
    \item The construction should be stable under perturbations.
\end{itemize}
The latter condition ensures the appropriate reduction, as discussed in section \ref{sec:reduction}.
If these conditions are met, we can further try to extend $\cL$ to $\cM'\supset\cM$ as defined in equation (\ref{qmPrime}).

The idea to look at leaves arises from an important result in the theory of Poisson manifolds: every Poisson manifold decomposes into a foliation of symplectic leaves.
This is proven via the Frobenius theorem, which relates smooth involutive distributions of constant rank in the tangent bundle of a manifold with foliations of the same \cite{Michor_2008}.\\
This specific result does not apply to the present context of a degenerate symplectic form, but we can copy the idea and define suitable distributions\footnote{Here, we do not necessarily demand involutivity, smoothness or even constant rank.} in the tangent bundle of $\mathcal{M}$ which hopefully result in approximate leaves in $\mathcal{M}$, which in turn can be studied numerically.

\subsection{The Hybrid Leaf}
\label{hybridLeaf}

While there are multiple approaches to define foliations of $\mathcal{M}$, examples show that the so-called \textit{hybrid leaf} is the most promising one, especially from a numerical point of view \cite{Felder_2023}. It comes in two flavors as we will see.

For a given point $x\subset\Tilde{\mathbb{R}}^D$ we put $\theta=(\theta^{ab}(x))$ and look at its (imaginary) eigenvalues
$\lambda_a$, ordered such that $\vert\lambda_a\vert\geq\vert\lambda_b\vert$ when $a<b$, coming with the (complex) eigenvectors $v_a$.
We can thus assume $\lambda_{2s-1}=+i\phi_s$ and $\lambda_{2s}=-i\phi_s$ for $s=1,\dots,\operatorname{rank}(\theta)/2$ and $\phi_s=\vert \lambda_{2s}\vert$ and $\lambda_a=0$ for $a>\operatorname{rank}(\theta)$.
We define $l$ as the largest index $a$ such that $\lambda_a\geq C$ where $C$ again is a small cutoff, thus redefining the effective dimension.

Now, the vectors $w_{2s-1}:=\operatorname{Re}(v_{2s-1})=\operatorname{Re}(v_{2s})$ and $w_{2s}:=\pm\operatorname{Im}(v_{2s-1})=\mp\operatorname{Im}(v_{2s})$ for $s=1,\dots,l/2$ span the combined eigenspaces corresponding to $\lambda_{2s-1}$ and $\lambda_{2s}$.
Thus, we define $V_x:=\langle w_1,\dots,w_l \rangle$ as a representative of the distribution in target space. Via push-forward to $\mathcal{M}$, we thus define the distribution $(Tq)(\sqcup_{x\in\widehat{\mathbb{R}}^D}V_x)\subset T\mathcal{M}$. 
\\
Note that we do not know if the rank of the distribution is constant, or if it is exactly integrable.
Thus approximate leaves, for which we write $\cL$ , have to be determined numerically.

While this leaf is associated
to target space (as $\theta^{ab}$ is), we can also construct a 
similar leaf which is more directly associated to the quantum manifold, by replacing $\theta$ with the would-be symplectic form $\omega:=(\omega_{ab}(x))$. The resulting leaf $\cL$ will be called \textit{hybrid leaf using} $\omega$. It has the advantage that  the resulting would-be symplectic form $\omega_\mathcal{L}$ is nondegenerate by construction.

\subsection{The (would-be) Quantization Map}
\label{lim}

Assume that we found an approximate leaf $\mathcal{L}\subset \mathcal{M}$. We can then refine the preliminary quantization map (\ref{quantmapCS}) as follows
\begin{align}
    \label{quantmapCSL}
    \cQ:\quad
    &\mathcal{C}(\mathcal{L})\to\operatorname{End}(\mathcal{H})\nonumber\\
    &\phi\mapsto \frac{\alpha}{(2\pi)^{k/2}}\int_\mathcal{L}\Omega_\mathcal{L} \,\phi\ket{\cdot}\bra{\cdot} \ ,
\end{align}
where $\Omega_\mathcal{L}:=\frac{1}{(l/2)!}\omega_\mathcal{L}^{\wedge l/2}$. We also refine the defining relation for $\alpha$ as $\frac{\alpha}{(2\pi)^{l/2}}V_\omega=N$ with $V_\omega:=\int_\mathcal{L}\Omega_\mathcal{L}$.
Of course,  the symbol map can also be restricted to a map
\begin{align}
\operatorname{Symb}:\quad &\operatorname{End}(\mathcal{H})\to\mathcal{C}(\mathcal{L})\nonumber\\
    &\Phi\mapsto \bra{\cdot}\Phi\ket{\cdot}
\end{align}
by restricting each function in the image to $\mathcal{L}$.

In order to be an acceptable quantization map, $\cQ$ should satisfy the axioms from section \ref{QS}.
Earlier results suggest that these are fulfilled approximately if
\begin{align}
    \label{Hypothesis}
    \quad \cQ(1_\mathcal{M})\approx \mathbb{1}_\mathcal{H}\ ,\quad n_X \cQ(\mathbf{x}^a)\approx X^a \ , \quad \{\mathbf{x}^a,\mathbf{x}^b\}\approx \theta^{ab}
\end{align}
and furthermore the matrix configuration is \textit{almost local} (see \cite{Steinacker_2021} for the definition). Here, $n_X$ is a proportionality constant and $\{,\}$ is the Poisson bracket induced by $\omega_\mathcal{L}$.
\\
Linearity and axiom two are satisfied by construction. (\ref{Hypothesis}.1) approximately implies axiom two, (\ref{Hypothesis}.2) approximately implies axiom four. If we pass to almost local quantum spaces, (\ref{Hypothesis}.3) supports axiom three \cite{Steinacker_2021, Felder_2023}.\\
Furthermore, (\ref{Hypothesis}.3) clarifies the question whether the Poisson structures induced by $\omega_\mathcal{L}$ or $\theta^{ab}$ should be used, by demanding that they approximately agree.

We thus propose to consider $(\mathcal{L},\omega_\mathcal{L})$ together with $\cQ:\mathcal{C}(\mathcal{L})\to\operatorname{End}(\mathcal{H})$ and $\mathbf{x}:\mathcal{L}\to\mathbb{R}^D$ as a semi-classical limit for a given matrix configuration $X^a$, while the conditions (\ref{Hypothesis}) have to be verified numerically. Algorithms for that purpose can be found in \cite{Felder_2023}. These were implemented in the Mathematica package \textit{QMGsurvey} \cite{Felder_2023_Software}.

\section{Examples}
\label{EXAMPLES}

To demonstrate the construction from section \ref{QMG} and \ref{LEAVES}, we provide a number of exemplary matrix configurations which are investigated numerically, using the Mathematica package\footnote{The relevant notebooks can also be found there.} \textit{QMGsurvey} \cite{Felder_2023_Software}. The implemented algorithms are described in \cite{Felder_2023}.

Perhaps the most canonical examples are provided by so called \textit{quantized coadjoint orbits} of compact semi-simple Lie groups. For these, one can prove rigorously that already the construction described in section \ref{limPrelim} provides a proper semi-classical limit. This makes the discussion in section \ref{LEAVES} unnecessary, and one immediately finds $\mathcal{L}=\mathcal{M}$ \cite{Steinacker_2021,Felder_2023,Bernatska_2012,Neeb_1994,Kostant_1982}.

Next, we consider examples where the construction of section \ref{LEAVES} is  necessary, including the \textit{squashed fuzzy sphere} (section \ref{sfs}), the \textit{perturbed fuzzy sphere} (section \ref{pfs}), the \textit{fuzzy torus} (section \ref{ft}) and  \textit{fuzzy} $\mathbb{C}P^2$ (section \ref{sfcp2}).

\subsection{The Squashed Fuzzy Sphere}
\label{sfs}

The matrix configuration of the \textit{(round) fuzzy sphere} is the most basic example of a coadjoint orbit, based on the Lie group $SU(2)$.
For given $N\in\mathbb{N}$, we define three hermitian matrices $J^a$ (thus $D=3$) as the orthonormal Lie group generators of $SU(2)$ acting on an orthonormal basis of the $N$ dimensional irreducible representation.\\
Using the quadratic Casimir $C_N^2=\frac{N^2-1}{4}$, we define the three matrix configuration\footnote{These then satisfy $\sum_a X^aX^a=\mathbb{1}$ in analogy to the Cartesian embedding functions of the ordinary sphere.}
\begin{align}
    X^a=\frac{1}{\sqrt{C_N^2}}J^a \ ,
\end{align}
which constitutes the round fuzzy sphere \cite{Madore_1992,hoppe1982QuaTheMasRelSurTwoBouStaPro}.
It is  straightforward to calculate the quasi-coherent states and the hermitian form using the $SU(2)$ symmetry. It is then clear that for $N>1$ the dimension of the quantum manifold is two and the (would-be) symplectic form is nondegenerate, thus there is no need to look at leaves. It can be rigorously proven that $\cQ$ is a quantization map and all assertions in (\ref{Hypothesis}) hold. This generalizes to more generic coadjoint orbits, cf.  \cite{Steinacker_2021,Felder_2023}.

The simplest nontrivial matrix configuration derived from the round fuzzy sphere is the \textit{squashed fuzzy sphere} with squashing parameter $\alpha\geq0$. It is defined through the matrix configuration $(\Bar{X}^a):=(X^1,X^2,\alpha X^3)$. Via perturbation theory in the parameter $\alpha$ it can be proven that for $N>2$ and $0<\alpha<1$ the dimension of the quantum manifold is three, implying that the (would-be) symplectic form is degenerate. This behavior is called \textit{oxidation}, and can in fact be visualized \cite{Felder_2023}.
Thus here, we look for two dimensional leaves of the quantum manifold.

\begin{figure}[ht]
\centering
\begin{minipage}{.4\textwidth}
  \centering
  \includegraphics[height=.7\linewidth]{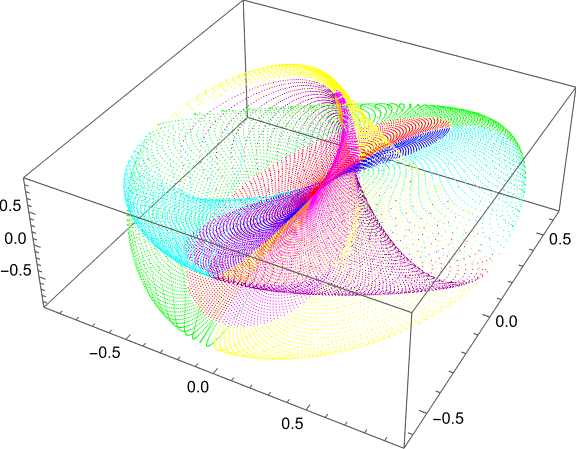}
\end{minipage}%
\begin{minipage}{.4\textwidth}
  \centering
  \includegraphics[height=.7\linewidth]{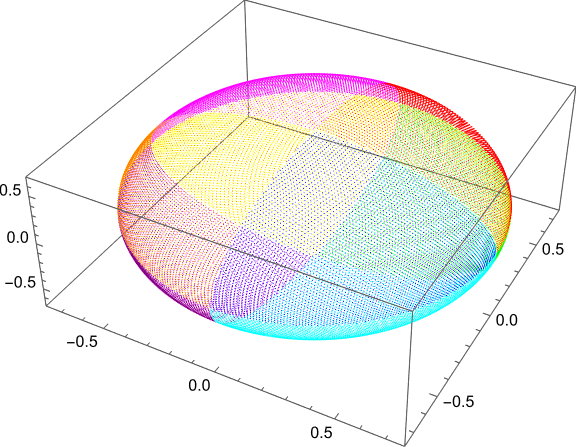}
\end{minipage}%
\caption{A covering with local coordinates of the hybrid leaf through $(1,2,1)\in\mathbb{R}^3$ for $N=4$, $\alpha=0.9$. Left: quantum manifold $\mathcal{M}$, right: embedded quantum space $\Tilde{\mathcal{M}}$}
\label{fig:sfsSpaces}
\end{figure}

In figure \ref{fig:sfsSpaces} we see a plot of a numerically constructed covering with local coordinates of the hybrid leaf. At least as far as visible, the distribution appears to be perfectly integrable.

\begin{figure}[ht]
\centering
\begin{minipage}{.4\textwidth}
  \centering
  \includegraphics[height=.6\linewidth]{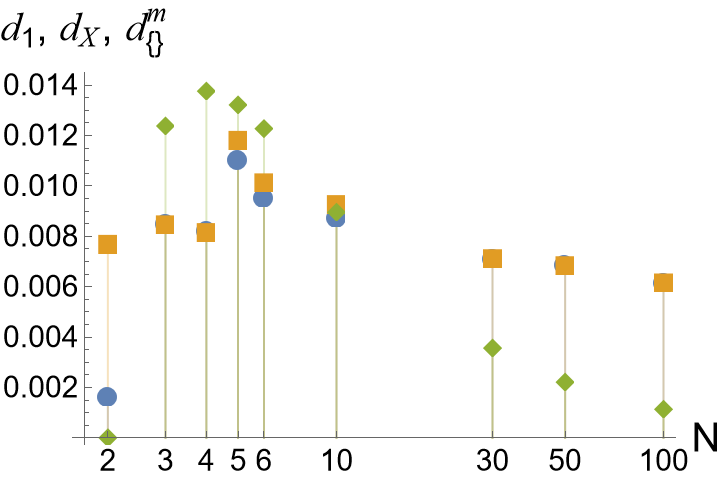}\\
  \includegraphics[height=.6\linewidth]{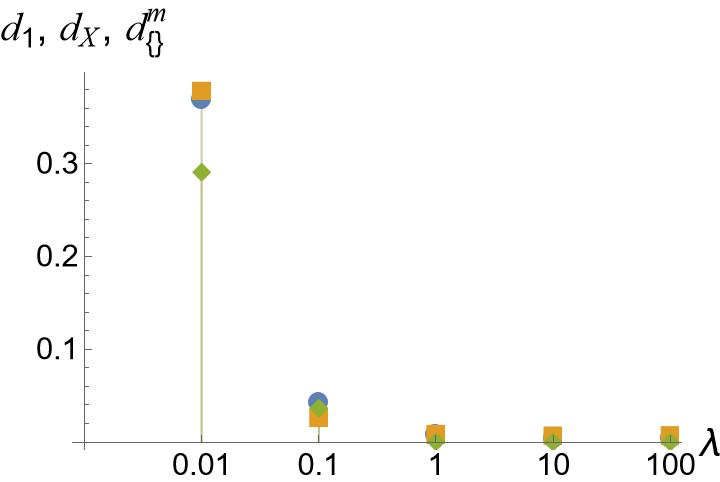}
\end{minipage}%
\begin{minipage}{.4\textwidth}
  \centering
  \includegraphics[height=.6\linewidth]{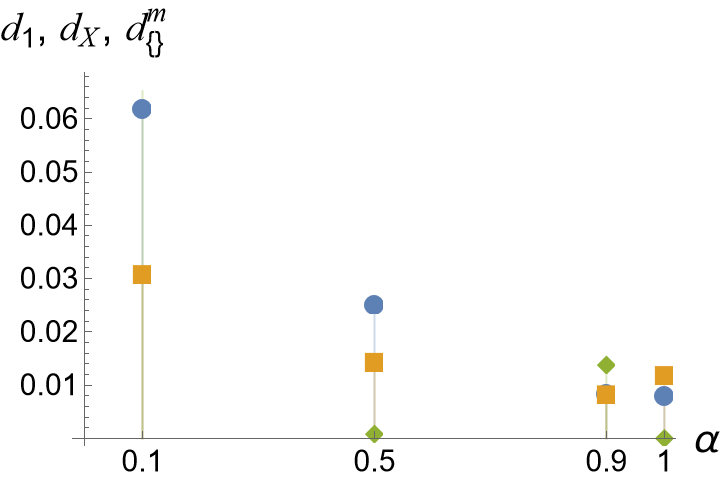}\\
  \includegraphics[height=.6\linewidth]{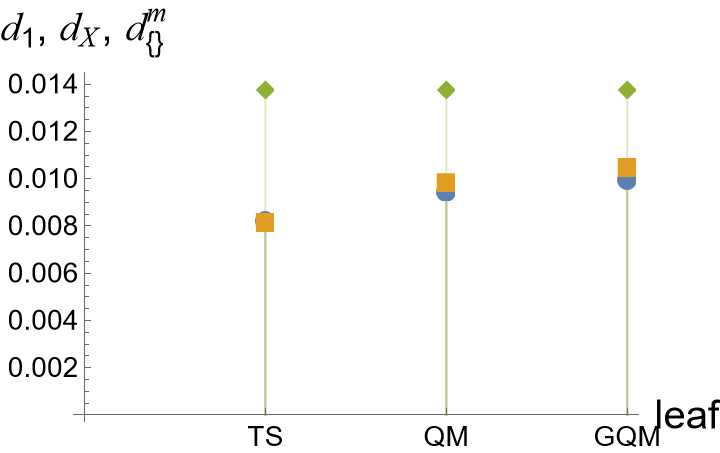}
\end{minipage}%
\caption{Modified\footref{jointfootnote} relative deviation of (\ref{Hypothesis}.1) ($d_\mathbb{1}$, blue), (\ref{Hypothesis}.2) ($d_X$, orange) and (\ref{Hypothesis}.3) ($d^m_{\{\}}$, green). The generic parameter choice is the setup of figure \ref{fig:sfsSpaces}. Top left: dependence on $N$ ($m=100$), top right: dependence on $\alpha$ ($m=100$), bottom left: dependence on the choice of leaf ($m=2$) with $\lambda\cdot(1,2,1)\in\mathbb{R}^3$ lying in and hence specifying the leaf, bottom right: dependence on the choice of distribution ($m=100$) with TS being the hybrid leaf, QM being the hybrid leaf using omega and GQM being the same with differently constructed coordinates}
\label{fig:sfsValidation}
\end{figure}

In order to verify how well the conditions in (\ref{Hypothesis}) are met, in figure \ref{fig:sfsValidation}
the modified\footnote{This is $d_\mathbb{1}:=\frac{\Vert \cQ(1_\mathcal{M})-\mathbb{1}_\mathcal{H} \Vert_\infty}{\sqrt{N}\Vert \mathbb{1}_\mathcal{H}\Vert_\infty}$, $d_X:=\frac{\Vert (n_X \cQ(\mathbf{x}^a)-X^a) \Vert_\infty}{\sqrt{N}\Vert (X^a)\Vert_\infty}$ respectively $d^m_{\{\}}:=m\cdot\frac{\Vert (\{\mathbf{x}^a,\mathbf{x}^b\}-\theta^{ab})) \Vert_\infty}{\Vert (\theta^{ab})\Vert_\infty}$ where $n_X$ is chosen such that $\Vert \cdot\Vert_\infty$ applied to (\ref{Hypothesis}.2) holds true and we think of the equations as tensorial equations. $m$ is only included for appropriate visualization and $\sqrt{N}$ is needed for scaling reasons \cite{Felder_2023}.\label{jointfootnote}} 
relative deviations of the equations are shown, depending on various choices.
First, one notes that (\ref{Hypothesis}.3) is of much better quality than (\ref{Hypothesis}.1) and (\ref{Hypothesis}.2) (in fact, it is \textit{rescaled} according to $m$ in the plots). This has the simple reason that the calculations are pointwise here, while for the other two quantities numerical integration over the leaf is necessary,  causing numerical errors.\\
Further, one sees that  the quality improves for $N\to\infty$, $\alpha\to 1$ and $\vert x\vert\to\infty$, while the conditions (\ref{Hypothesis}) always hold to a satisfactory extent. The choice of distribution is not significant here.

\subsection{The Perturbed Fuzzy Sphere}
\label{pfs}

Another generalization of the round fuzzy sphere is the \textit{perturbed fuzzy sphere}.\\
While for the ordinary sphere $S^2$ the space of functions decomposes as $\mathcal{C}(S^2)\cong \bigoplus_{l=0}^\infty (l)$, for the fuzzy sphere the space of modes is truncated $\operatorname{End}(\mathcal{H})\cong \bigoplus_{l=0}^{N-1} (l)$ where in both cases $(l)$ stands for the $2l+1$ dimensional irreducible representation of $SU(2)$, spanned by the classical spherical harmonics $Y_l^m$ and the fuzzy spherical harmonics $\hat{Y}_l^m$, respectively.\\
Given a \textit{cutoff} $0\leq c\leq N-1$, we randomly\footnote{We take $\hat{Y}_l^m$ for $l\leq c$ as a basis and pick evenly distributed coefficients between $[-1,1]$.} choose three hermitian elements $\cA^a$ in the span of $\hat{Y}^l_m$ for $l\leq c$. Given  these, we consider the matrix configuration 
$\Tilde{X}^a = X^a+\cA^a$, which defines the perturbed fuzzy sphere of \textit{degree} $c$.

On the classical side, we can do the same by defining three real valued elements $a^a$ in the span of $Y^l_m$ for $l\leq c$ with the same coefficients and add these to the Cartesian embedding functions $\Tilde{x}^a:=x^a+a^a$. The image $(\Tilde{x}^a(S^2))$ then defines a perturbed sphere embedded into $\mathbb{R}^3$.

The matrices $A^a$ acquire the interpretation of \textit{gauge fields} on the quantum space $X^a$ \cite{Steinacker_2015}. Further, the matrix configuration $\Tilde{X}^a$ is almost local for $c\leq \sqrt{N}$,  which is the scale of noncommutativity (NC) on the fuzzy sphere $S^2_N$.
We therefore expect to obtain a good quality of the quantization map below this scale, and bad quality above \cite{Steinacker_2021}.

\begin{figure}[ht]
\centering
\begin{minipage}{.4\textwidth}
  \centering
  \includegraphics[height=.7\linewidth]{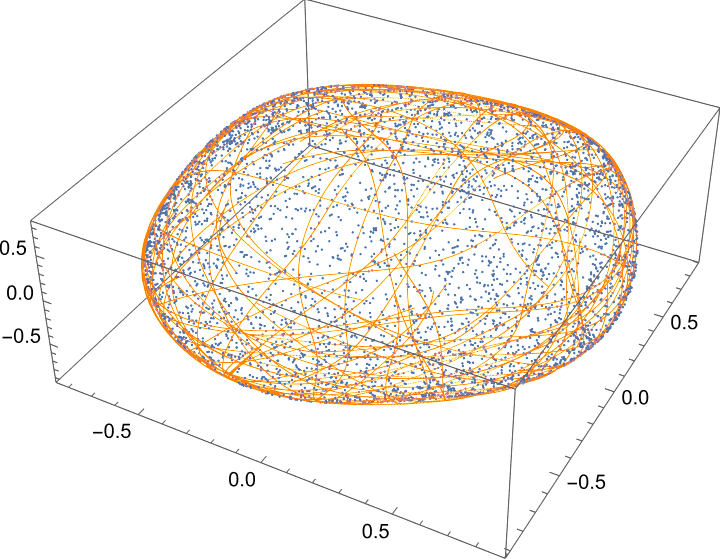}
\end{minipage}%
\begin{minipage}{.4\textwidth}
  \centering
  \includegraphics[height=.6\linewidth]{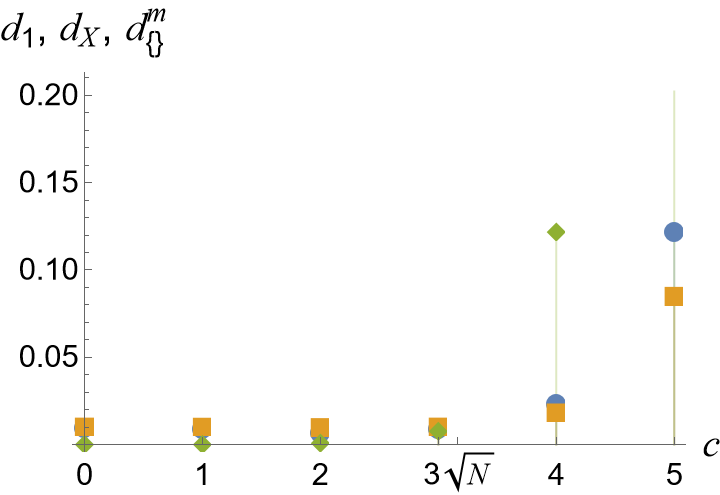}
\end{minipage}%
\caption{Left: comparison of the classical and quantum space constructed with identical random coefficients, blue: classical space -- image of random points on the sphere, orange: embedded quantum space -- scan of the hybrid leaf through $(1,2,1)\in\mathbb{R}^3$ for $N=10$ and $c=4$, right: plot similar to those of figure \ref{fig:sfsValidation} depending on $c$ ($m=2$)}
\label{fig:pfs}
\end{figure}

In figure \ref{fig:pfs} we see that the semi-classical limit of the quantum space defined through the hybrid leaf almost perfectly agrees with the classical space. Further, the expected dependence on the cutoff scale is confirmed: for $N=10$ we have $\sqrt{N}\approx 3.2$ and clearly the quality is rather good for $c$ below this scale, while it becomes significantly worse  above the NC scale.

\subsection{The Fuzzy Torus}
\label{ft}

The \textit{fuzzy torus} is an elementary example of a matrix configuration that is \textit{not} derived from a quantized coadjoint orbit.\\
For $N>1$, we define $q:=\exp{(2\pi i\frac{1}{N})}\in \mathbb{C}$ and two matrices $U$ and $V$ via $U_{ij}=\delta_{i,j+1}$ respectively $V_{ij}=\delta_{ij}q^{i-1}$. These satisfy the \textit{clock and shift algebra}
\begin{align}
    q^N=1,\quad U^N=\mathbb{1}=V^N,\quad U\cdot V=qV\cdot U.
\end{align}
It is a simple task to see that for $N\to\infty$ this reproduces the algebra $\mathcal{C}(T^2)$ for the classical torus $T^2$ \cite{Schneiderbauer_2016}.

With that, we set $X^1:=\frac{1}{2}(U+U^\dagger)$, $X^2:=\frac{1}{2i}(U-U^\dagger)$, $X^3:=\frac{1}{2}(V+V^\dagger)$ and $X^4:=\frac{1}{2i}(V-V^\dagger)$, constituting the matrix configuration $(X^a)$ of the fuzzy torus.\\
Numerically, for $N>2$ one finds that the maximal rank is $k=3$ while the effective dimension is  $l=2$. It is  hence unavoidable to consider leaves of $\mathcal{M}$.

\begin{figure}[ht]
\centering
\begin{minipage}{.4\textwidth}
  \centering
  \includegraphics[height=.7\linewidth]{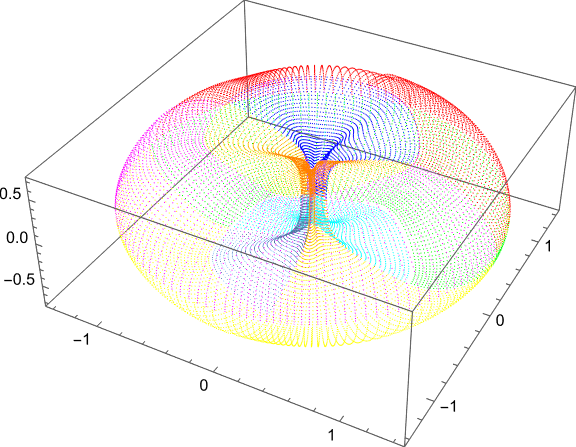}
\end{minipage}%
\begin{minipage}{.4\textwidth}
  \centering
  \includegraphics[height=.6\linewidth]{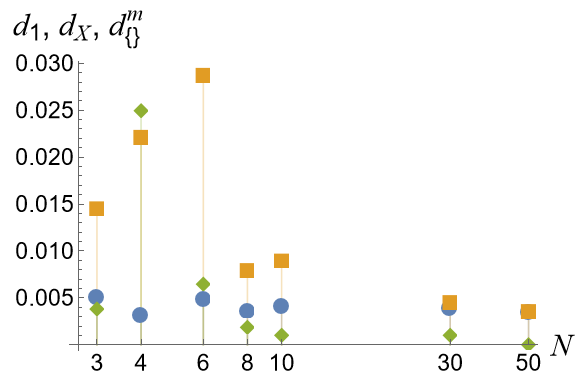}
\end{minipage}%
\caption{Left: a covering with local coordinates of the hybrid leaf through $(1,2,1,0)\in\mathbb{R}^3$ for $N=5$ -- a stereographic plot of the embedded quantum space, right: plot similar to those of figure \ref{fig:sfsValidation} depending on $N$ ($m=0.02$)}
\label{fig:tor}
\end{figure}

In figure \ref{fig:tor} we can see that the semi-classical limit defined through the hybrid leaf
reproduces a torus to surprising accuracy, already for $N=5$. Further, we find that for $N\to\infty$ the quality of the quantization map improves further, while the conditions of (\ref{Hypothesis}) hold to a good extent.
This is an important result, since we cannot rely here on group theory to determine $\mathcal{M}$ or $\cQ$,  even in the unperturbed case.

\subsection[The Squashed Fuzzy CP2]{The Squashed Fuzzy $\mathbb{C}P^2$}
\label{sfcp2}

The \textit{fuzzy} $\mathbb{C}P^2$ is a natural 4-dimensional  generalization of the fuzzy sphere, replacing  $SU(2)$ by $SU(3)$.
For given $n\in \mathbb{N}$, we define eight hermitian matrices $T^a$ (thus $D=8$) as the orthonormal Lie group generators of $SU(3)$ acting on an orthonormal basis of the irreducible representation\footnote{Numerically, they can be constructed via a Mathematica package \cite{Shurtleff_2009}.} $\mathcal{H}_n = (n,0)$. Then the dimension of the underlying Hilbert space is  $\dim \mathcal{H}_n = N=\frac{(n+1)(n+2)}{2}$.
Using the quadratic Casimir $C^2_n=\frac{1}{3}(n^2+3n)$, we define the eight matrices
\begin{align}
    X^a=\frac{1}{\sqrt{C^2_n}}T^a \ ,
\end{align}
constituting the fuzzy $\mathbb{C}P^2$ $(X^a)$.
Since this is  again a quantized coadjoint orbit, everything works out perfectly and one finds $\mathcal{L}=\mathcal{M}\cong \mathbb{C}P^2$ (thus $k=l=4$) \cite{Steinacker_2021, Felder_2023}.

\begin{figure}[ht]
\centering
\begin{minipage}{.4\textwidth}
  \centering
  \includegraphics[height=.7\linewidth]{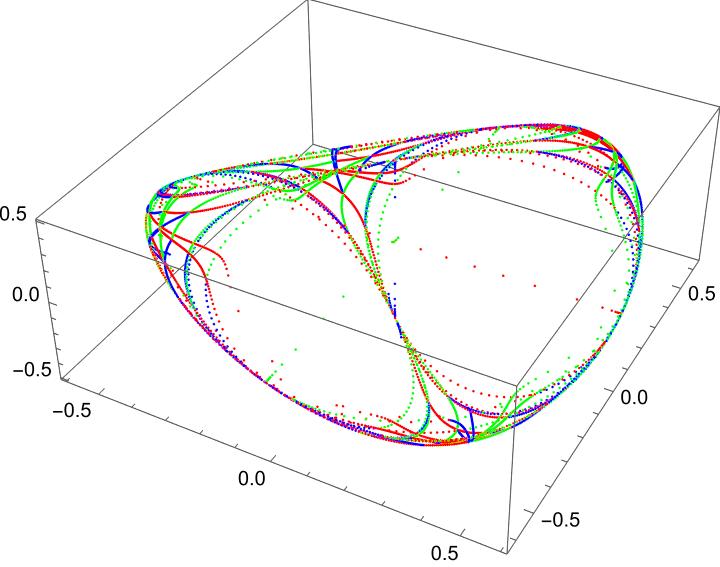}
\end{minipage}%
\begin{minipage}{.4\textwidth}
  \centering
  \includegraphics[height=.7\linewidth]{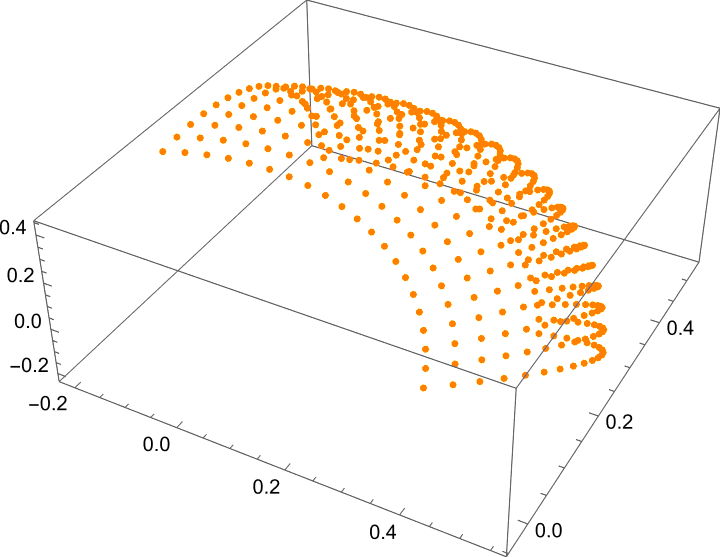}
\end{minipage}%
\caption{Left: embedded quantum space for $n=3$ -- image of Cartesian coordinate lines in target space, right: embedded quantum space for $n=3$ (different perspective) -- two directions of local coordinates in the hybrid leaf}
\label{fig:fcp2}
\end{figure}

As for the fuzzy sphere, we can squash  fuzzy $\mathbb{C}P^2$ by multiplying the matrices $X^a$ with some parameters $\alpha_a$ to obtain the matrix configuration $\Bar{X}^a:=\alpha_a X^a$, which defines the squashed fuzzy $\mathbb{C}P^2$. We will focus on the case  $\alpha_3=\alpha_8=\alpha\in[0,1]$  (corresponding to the Cartan generators),  with the remaining $\alpha_a = 1$, respectively where we (randomly) choose $\alpha_a$ within $[0,1]$.
\\
Using numerical simulations (for moderate squashing), one finds the maximal rank and the effective dimension to be $k=8$ and $l=4$, reflecting again the oxidation. Therefore the use of foliations is indispensable in order to obtain a semi-classical limit which is stable under perturbation.
Unfortunately, due to the increasing numerical expenses, it was not possible to generate a satisfactory covering with local coordinates of the hybrid leaf with the available computational capacity. Thus, we cannot provide significant data on the quality of the semi-classical limit, but we conjecture that the behavior is very similar to the squashed fuzzy sphere.\\
What can be done is to look\footnote{Note that due to $D>3$ some projection has to be performed, meaning that we can and must look at the geometries from different perspectives. See \cite{Felder_2023} for a thorough discussion.} at the image of Cartesian (or other) coordinate lines in target space in order to obtain a global picture of $\mathcal{M}$. Further, the hybrid leaf can be investigated locally by constructing local coordinates. Both is done in figure \ref{fig:fcp2} and we see that in principle the construction still works splendidly.

\section{Conclusion}

In this paper, we give a refined definition of a quantum manifold $\mathcal{M}$ which can be associated to any given matrix configuration via  quasi-coherent states, generalizing the definition in \cite{Steinacker_2021}. In appendix \ref{appendix:quantummanifold}, this is shown  to provide an immersed submanifold in $\mathbb{C}P^{N-1}$.
This quantum manifold is endowed with a hermitian form, which defines the quantum metric and the (would-be) symplectic form. This allows to define a preliminary quantization map, which also induces (would-be) Cartesian embedding functions, thus providing the basic ingredients for a semi-classical description of the underlying matrix configuration.\\
While that construction is perfectly satisfactory e.g. for quantized coadjoint orbits, the construction suffers from oxidation in the general case, leading to a quantum manifold with too many dimensions.
We propose some approaches to reduce this quantum manifold to an underlying minimal or irreducible core.
In particular, we propose to 
consider foliations of the quantum manifold, and a restriction to a suitably chosen leaf.
\\
We also study numerically several examples of deformed matrix configurations, where such a restriction is seen to be essential. In particular, one approach to achieve the desired reduction (denoted as \textit{hybrid leaf approach}) is found to be quite satisfactory, at least numerically. It is both computationally efficient and conceptually appealing, as it turns out to be stable under perturbation;
in particular, the variant based on foliations defined by $\omega$ leads by construction to a symplectic form.

Explicit calculations strongly suggest that (at least for large $N$), the hybrid leaf construction leading to $(\mathcal{L},\omega_\mathcal{L})$ together with the maps $\cQ:\mathcal{C}(\mathcal{L})\to\operatorname{End}(\mathcal{H})$ and $\mathbf{x}:\mathcal{L}\to\mathbb{R}^D$ (approximately) defines a satisfactory semi-classical limit for the given almost local matrix configurations, satisfying the requirements (\ref{Hypothesis}). This has been tested for examples derived from quantized coadjoint orbits (the squashed and perturbed fuzzy spheres), but also  an example where this is not the case (the fuzzy torus), using the Mathematica package \textit{QMGsurvey} \cite{Felder_2023_Software}.
Although for the higher dimensional squashed fuzzy $\mathbb{C}P^2$ the computational demand has been too high to accurately calculate the quantization map, local coordinates in the hybrid leaf can still be constructed.
\\
We also produce several visualizations for the perturbed fuzzy sphere, which nicely exhibit a meaningful semi-classical limit. The  examples also support the conjecture of \cite{Felder_2023} that the semi-classical limit is of good quality if the cutoff in the $SU(2)$ modes is chosen below the scale of noncommutativity  $\sqrt{N}$, but not beyond.
\\
While these results are very encouraging, there are many open problems.
 Even though the hybrid leaf approach appears to be very promising,  it is not yet clear how to properly choose a leaf of the foliation. One may also hope to find other, perhaps iterative procedures which effectively reduce some given matrix configuration to its ``irreducible core'', thus reversing the oxidation of the underlying quantized symplectic space through perturbations. In a sense, this problem can be viewed as a noncommutative version of finding optimal Darboux-like coordinates for some given 
almost-commutative  matrix configuration.
\\
Finally, results on the integrability and smoothness of the defining distributions would be very desirable.
\\
The present problem and the methods under discussion should be of considerable interest for numerical simulations of large-$N$ matrix models related to string theory, see
e.g. \cite{Nishimura:2022alt,Nishimura:2011xy}. In particular, they should allow to understand the geometrical meaning of the dominant matrix configurations of such matrix models, and to assess their interpretation in terms of  space or space-time.

\paragraph{Acknowledgments}

This research was funded in whole or in part by the Austrian Science Fund (FWF) 10.55776/P32086.

\printbibliography

\appendix

\section{Analyticity of the Eigensystem}
\label{appendix:analyticity}

Smooth dependence of the quasi-coherent states on $x\in\Tilde{\mathbb{R}}^D$ is an essential property of the whole construction. For that reason, important results on the analytic and especially smooth dependence of the eigensystem on multiple parameters of a hermitian matrix are summarised here.

\textbf{Theorem:} Let $A:U\subset \mathbb{R}^n\to\operatorname{Herm}(\mathcal{H})$ be an analytic function from an open subset $U$ of $\mathbb{R}^n$ into the set of hermitian operators on a finite dimensional Hilbert space $\mathcal{H}$. Let $x\in U$ and $\lambda_x$ be an eigenvalue of $A(x)$ of multiplicity $1$ with corresponding eigenvector $v_x$.\\
Then, there exists an open neighborhood $V\subset U$ of $x$ and analytic functions $\lambda:V\to \mathbb{R}$ and $v:V\to\mathcal{H}$ such that $\lambda(x)=\lambda_x$, $v(x)=v_x$ with $\lambda(y)$ being an eigenvalue of $A(y)$ of multiplicity $1$ with corresponding eigenvector $v(y)$ $\forall y\in V$.

This follows from \cite{Kurdyka_2006} (section 7), generalizing results by Rellich and Kato for a single parameter \cite{Rellich_1969,Kato_1995}.

The Hamiltonian $H:\mathbb{R}^D\to\operatorname{Herm}(\mathcal{H})$ from equation (\ref{Hamiltonian}) obviously is analytic. Thus we can apply the theorem to all points in $\Tilde{\mathbb{R}}^D:=\{x\in\mathbb{R}^D\vert \operatorname{dim}(E_x)=1\}$ (with $E_x$ being defined as the eigenspace of $H_x$ corresponding to the smallest eigenvalue $\lambda_{x}$).\\
Especially, this implies that $\Tilde{\mathbb{R}}^D$ is open in $\mathbb{R}^D$, the map $\lambda:\Tilde{\mathbb{R}}^D\to\mathbb{R}$ (assigning the lowest eigenvalue of the Hamiltonian) is smooth and we can locally choose the quasi-coherent states (the corresponding normalized eigenstates) in a smooth way.

\section{The Quantum Manifold}
\label{appendix:quantummanifold}

In section \ref{mfp} the objects $k$, $\widehat{\mathbb{R}}^D$, $\mathcal{M}$ and $q$ have been defined.
By construction we deal with a smooth surjection $q:\widehat{\mathbb{R}}^D\to\mathcal{M}\subset\mathbb{C}P^{N-1}$ of constant rank $k$.\\
Therefore, we recall the constant rank theorem:
\begin{quote}
\textbf{Theorem:} \textit{``Suppose $M$ and $N$ are smooth manifolds of dimensions $m$ and $n$, respectively, and $F:M\to N$ is a smooth map with constant rank $r$. For each $p\in M$ there exist smooth charts $(U,\phi)$ for $M$ centered at $p$ and $(V,\psi)$ for $N$ centered at $F(p)$ such that $F(U)\subseteq V$, in which $F$ has a coordinate representation} [$\hat{F}=\psi\circ F\circ \phi^{-1}:\phi(U)\to \psi(V)$] \textit{of the form $\hat{F}(x^1,\dots,x^r,x^{r+1},\dots,x^m)=(x^1,\dots,x^r,0,\dots,0)$.''} -- theorem 4.12 in \cite{Lee_2003}.
\end{quote}

For any $x\in\widehat{\mathbb{R}}^D$ the theorem guarantees the existence of corresponding charts $(U,\phi)$ for $\widehat{\mathbb{R}}^D$ and $(V,\psi)$ for $\mathcal{M}$.
We will use these to define a chart for $\mathcal{M}$, but before we can do so we need two technical lemmas.

In the first lemma we construct a \textit{prototypical chart} $(\mathcal{U},\bar{\beta}^{-1})$ around $q(x)\in\mathcal{M}$, using the constant rank theorem.

\textbf{Lemma 1:}
Consider $q:\widehat{\mathbb{R}}^D\to\mathcal{M}\subset\mathbb{C}P^{N-1}$ as in section \ref{mfp}. Then for each $x\in\widehat{\mathbb{R}}^D$ there exists an open subset $\Bar{U}\subset \mathbb{R}^k$ and a subset $\mathcal{U}\subset \mathcal{M}$ around $0$ respectively $q(x)$ and a smooth bijection $\Bar{\beta}:\Bar{U}\to\mathcal{U}$ between them.

\textbf{Proof:}
We consider the charts and the map $q':=\psi\circ q\circ \phi^{-1}: \phi(U)\to\psi(V)$, coming from the constant rank theorem.\\
Now, we define the sets $U':=\phi(U)$, $\mathcal{U}:=q(U)$ and the smooth map $\beta':=q\circ \phi^{-1}:U'\to \mathcal{U}$, with $\beta'$ being bijective by construction.\\
Further, $\beta'=\psi^{-1}\circ q'$ and since $\psi$ is a diffeomorphism, $\beta'$ still has constant rank $k$ and depends only on the first $k$ coordinates $\chi^1,\dots,\chi^k$ in $U'\subset\mathbb{R}^D$.\\
We define\footnote{Here, we implicitly identified $\mathbb{R}^k\cong \mathbb{R}^k\times\{0\}\subset\mathbb{R}^D$ and similarly for $\mathbb{R}^{D-k}$.} $\Bar{U}:=U'\cap \mathbb{R}^k$ and $\Bar{\beta}:=\beta'\vert_{\Bar{U}}$.\\
By shrinking\footnote{Especially such that $\Bar{U}\times \mathbb{R}^{D-k}$ contains $U'$.} $U$ we can make $\Bar{\beta}$ surjective. Since $\Bar{\beta}$ has full rank it is an immersion, thus locally injective and we can shrink $U$ once again in order to make $\Bar{\beta}$ bijective. $\square$

A little comment on the notation may be helpful. All quantities marked with a \textit{prime} are related to the space $\mathbb{R}^D$ that is the target of the chart $\phi$, while the quantities marked with a \textit{bar} live on $\mathbb{R}^k\cong\mathbb{R}^k\times \{0\}\subset\mathbb{R}^D$.

Recall $\mathcal{N}_x$ from section \ref{NullSpace}.
We find that for all $y\in \bar{U}$ the restriction of $\phi^{-1}$ to $\mathcal{N}'_y:=(\{y\}\times\mathbb{R}^{D-k})\cap U'$ is a diffeomorphism from $\mathcal{N}'_y$ to $\mathcal{N}_{\phi^{-1}(y,0)}\cap U$.\\
To see this we note that $\phi^{-1}(\mathcal{N}'_y)$ lies within $\mathcal{N}_{\phi^{-1}(y,0)}\cap U$ as $\beta'$ is constant here. Further $q'$ is injective on $U'\cap\mathbb{R}^k$, thus only points in $\mathcal{N}'_y$ are mapped to $\mathcal{N}_{\phi^{-1}(y,0)}\cap U$.

For any $z\in\widehat{\mathbb{R}}^D$ this immediately implies $\dim(T_z\mathcal{N}_z)=D-k$, while obviously $T_z\mathcal{N}_z\subset \operatorname{ker}(T_zq)$. Thus, $T_z\mathcal{N}_z=\operatorname{ker}(T_zq)$ what in turn implies that the whole $\mathcal{N}_z$ lies within $\widehat{\mathbb{R}}^D$.
\\
This leads us to the following lemma which shows that $\mathcal{N}_W:=\cup_{x\in W}\mathcal{N}_x$ is open in $\widehat{\mathbb{R}}^D$ if $W$ is.

\textbf{Lemma 2:}
Let $W\subset\widehat{\mathbb{R}}^D$. If $W$ is open in $\widehat{\mathbb{R}}^D$, then also $\mathcal{N}_W$ is open in $\widehat{\mathbb{R}}^D$.

\textbf{Proof:} Recall the setup from the proof of lemma 1. Since $\mathcal{N}_W\cup\mathcal{N}_X=\mathcal{N}_{W\cup X}$, it suffices to show the claim for $W\subset U$.
We define $W':=\phi(W)\subset U'$ what is clearly open. By the above we find $\mathcal{N}_{W}\cap U=\phi^{-1}((W'+\{0\}\times \mathbb{R}^{D-k})\cap U')$. Since $\phi$ is a diffeomorphism, this shows that $\mathcal{N}_{W}\cap U$ is open.\\
Consider now a point $y\in \mathcal{N}_W$. By construction there is a $y'\in W$ such that $y\in\mathcal{N}_{y'}$. Since $\mathcal{N}_{y'}$ is convex (see section \ref{NullSpace}), it contains the straight line segment that joins $y$ with $y'$.\\
Then, we can pick \textit{ordered} points $y_\alpha$ for $\alpha=0,\dots,n$ on this line segment with $y_0=y'$ and $y_n=y$ such that the corresponding $U_\alpha$ (which we get in the proof of lemma 1 for the point $y_\alpha$) cover the line segment (w.l.o.g. we have $U_\alpha\cap U_\beta\neq\emptyset$ iff $\alpha$ and $\beta$ are consecutive).
For $\alpha>0$ we inductively define $W_{\alpha}:=\mathcal{N}_{W_{\alpha-1}}\cap U_{\alpha-1}\cap U_{\alpha}\subset \mathcal{N}_W$ with $W_0:=W$.\\
Repeating the above discussion, all $W_\alpha$ are open. By construction, all $W_\alpha$ contain $\mathcal{N}_{y'}\cap U_{\alpha-1}\cap U_{\alpha}$ and are thus nonempty. So finally, $y\in \mathcal{N}_{ W_n}\cap U_n \subset\mathcal{N}_{W}$.\\
This means that every point $y\in\mathcal{N}_{W}$ has an open neighborhood in $\mathcal{N}_{W}$. $\square$

This result is crucial and allows us to prove that $\mathcal{M}$ is a smooth manifold.

\textbf{Theorem:}
$\mathcal{M}$ as defined in section \ref{mfp} is a smooth manifold of dimension $k$.

\textbf{Proof:}
By lemma 1 we can cover $\mathcal{M}$ with prototypical charts $(\mathcal{U}_\alpha,\Bar{\beta}_\alpha^{-1})$ around $q(x_\alpha)$ for appropriate points $x_\alpha$, defining a \textit{prototypical atlas}.\\
Consider now two charts with indices $\beta,\gamma$ such that $\mathcal{U}_{\beta \gamma}:=\mathcal{U}_\beta\cap \mathcal{U}_\gamma\neq \emptyset$.
That the corresponding chart changes are smooth is obvious, but it remains to show that the charts consistently define a topology on $\mathcal{M}$. Then, by lemma 1.35 in \cite{Lee_2003}, $\mathcal{M}$ is a smooth manifold of dimension $k$.\\
Especially, this means that we have to show that (w.l.o.g.) $\bar{U}_{\beta \gamma}:=\Bar{\beta}_\beta^{-1}(\mathcal{U}_{\beta\gamma})$ is open in $\mathbb{R}^k$.\\
Since $\bar{U}_{\beta\gamma}\subset \bar{U}_\beta$, we only have to show that $\bar{U}_{\beta\gamma}$ is open in the latter. This is equivalent to the statement that there is no sequence in $\bar{U}_\beta\setminus \bar{U}_{\beta\gamma}$ that converges against a point in $\bar{U}_{\beta\gamma}$.\\
Assume to the contrary that such a sequence $(y_a)$ exists. Since $\phi_\beta$ is a diffeomorphism, this defines the convergent sequence $(x_a):=(\phi_\beta^{-1}(y_a))$ in $\widehat{\mathbb{R}}^D$ with $\lim_{a\to \infty}x_a=:x'_\beta$. By assumption, $q(x'_\beta)\in \mathcal{U}_{\beta\gamma}$, thus, there is an $x'_\gamma\in U_\gamma$ with $q(x'_\beta)=q(x'_\gamma)$. But then by definition, $x'_\beta\in \mathcal{N}_{x'_\gamma}$ and consequently $x'_\beta\in \mathcal{N}_{U_\gamma}$. Since the latter is open by lemma 2, this implies that some $x_{a_i}$ lie within $\mathcal{N}_{U_\gamma}$. But for these, we find $q(x_{a_i})\in \mathcal{U}_{\beta\gamma}$ and consequently $y_{a_i}\in \bar{U}_{\beta\gamma}$ what contradicts the assumption. $\square$

\textbf{Remark:}
As the rank of $q$ equals $k$ this shows that the inclusion $\mathcal{M}\hookrightarrow \mathbb{C}P^{N-1}$ is an immersion and thus $\mathcal{M}$ is an immersed submanifold of the latter.

If we identify $\mathbb{C}P^{N-1}\cong S^{2N-1}/U(1)$, where $S^{2N-1}\cong\{\ket{\psi}\in\mathbb{C}^N\vert\; \braket{\psi\vert \psi}=1\}$, we also get a natural identification of the tangent spaces\footnote{The tangent space of $S^{2N-1}$ at $\ket{\psi}$ is given by all vectors $\ket{v}\in\mathbb{C}^N$ that satisfy the constraint $\operatorname{Re}(\braket{\psi\vert v})=0$. In the quotient this means that the constraint has to be satisfied for all $U(1)\ket{\psi}$, meaning $\braket{\psi\vert v}=0$.} $T_{U(1)\ket{\psi}} \mathbb{C}P^{N-1}\cong \{\ket{v}\in\mathbb{C}^N\vert\; \braket{\psi\vert v}=0\}$.\\
On the other hand, complex projective space can be viewed as the natural fiber bundle $p:\mathbb{C}^N\setminus \{0\}\to\mathbb{C}P^{N-1}$, where in our identification the bundle projection $p$ maps $\ket{\psi}$ to $U(1)\ket{\psi}/\sqrt{\braket{\psi\vert \psi}}$, coming with the tangent map $T_{\ket{\psi}}p=(\mathbb{1}-\ket{\psi}\bra{\psi})$.\\
Then, we get an induced hermitian form $\hat{h}$ on $T\mathbb{C}P^{N-1}$ via $\hat{h}(U(1)\ket{\psi})(\ket{v},\ket{w})=\braket{v\vert w}$. By construction, its real part is proportional to the Fubini-Study metric and its imaginary part to the canonical symplectic form on $\mathbb{C}P^{N-1}$, the Kirillov-Kostant-Souriau symplectic form \cite{Kostant_1982}.\\
Since the inclusion $j:\mathcal{M}\hookrightarrow\mathbb{C}P^{N-1}$ is a smooth immersion, we can pull back $\hat{h}$ to $\mathcal{M}$ what we call $h_\mathcal{M}=g_\mathcal{M}+i\omega_\mathcal{M}$, providing us with a metric and an (in general degenerate but closed) would-be symplectic form. Especially, we find $h_{ab}(x):=(q^*h_\mathcal{M})(x)(\partial_a,\partial_b)= h_\mathcal{M}(q(x))(T_xq\cdot\partial_a,T_xq\cdot\partial_b)=h_\mathcal{M}(q(x))(T_{\kets{x}}p\cdot\partial_a\kets{x},T_{\kets{x}}p\cdot\partial_b\kets{x})=h_\mathcal{M}(q(x))(D_a\kets{x},D_b\kets{x})=\hat{h}(U(1)\kets{x})(D_a\kets{x},D_b\kets{x})$\\$=(D_a\kets{x})^\dagger D_b\kets{x}$, where we described $q=p\circ\kets{\cdot}$ via a local section and used $D_a\kets{x}=(\mathbb{1}-\kets{x}\bras{x})\partial_a\kets{x}=T_{\kets{x}}p\cdot\partial_a\kets{x}$.\\
But this means that the pullback of $h_\mathcal{M}$ along $q$ agrees with the $h_{ab}$ we have defined earlier, similarly for $g_\mathcal{M}$ and $\omega_\mathcal{M}$.

\end{document}